\documentstyle[epsf,aps,prl]{revtex} 
\begin{document}
\draft
\twocolumn[\hsize\textwidth\columnwidth\hsize\csname 
@twocolumnfalse\endcsname 
\title{ENERGY AND SYMMETRY OF ORDERED TWO-DIMENSIONAL DIPOLE CLUSTERS IN A
PARABOLIC CONFINEMENT}
\author{M. Golosovsky\footnote{e-mail golos@vms.huji.ac.il}*, Y. Saado, and
D. Davidov}
\address{The Racah Institute of Physics, The Hebrew University of
Jerusalem, Jerusalem 91904, Israel\\}
\date{\today} \maketitle
\begin{abstract} 
We report an experimental study of two-dimensional dipole clusters
formed by small magnetic particles floating at the air-water interface.  The
particles form hexagonally ordered clusters due to interplay between
dipole-dipole repulsion and confining external forces.  Optical images
of the clusters allow determination of area $S$, total energy
$E$, and the chemical potential $\mu$ for different particle numbers
$N<130$.  Asymptotic expressions for $E(N),\mu(N), S(N)$ account
fairly well for the smooth part of our experimental results.  However,
on top of the smooth dependences we observe
quasiperiodic fluctuations with dips at
``magic'' numbers corresponding to particularly symmetric cluster
configurations.  The magnitude and occurrence of these fluctuations
are related to the symmetry of the cluster and to the center of mass
position.  This is in full agreement with the recent
prediction of Koulakov and Shklovskii for Wigner crystal island.
\end{abstract}

\bigskip 
] 
\pagebreak

Confined two-dimensional (2D) systems of interacting particles, such as
electrons in quantum dots\cite{zhitenev}; vortices in plasma
\cite{plasma}, in superfluids\cite{superfluid} and in mesoscopic
superconductors\cite{Geim}; adatoms; colloidal crystals at liquid
interfaces\cite{pieranski}; electrons on the surface of the liquid
helium\cite{el on He}, have attracted much attention.  These systems
exhibit variety of interesting phenomena, including melting
\cite{pieranski,Skjeltorp,Zahn,bubeck,Seshadri,schweigert}, crystallization,
shell structure \cite{Bedanov,Lozovik} and ``magic''
numbers\cite{Martin}.  While quantum mechanics is certainly important
here, many features of these systems arise solely from geometry and
may be explained classically.  In particular, transport measurements
on a quantum dot when it forms Wigner crystal at low electron
concentration or in a strong magnetic field, reveal
periodic disappearance or merging of Coulomb blockade peaks
\cite{zhitenev}.  A
classical explanation of this puzzling phenomenon has been proposed by
Koulakov and Shklovskii \cite{Shklovskii} who traced it to the
intricate dependence of the crystal energy on its symmetry.  In
particular, Ref.\cite{Shklovskii} predicted that (i) the center of
mass position in the unit cell of the 2D Wigner crystal varies
periodically with the number of particles; (ii) the center of mass
sticks to positions of high symmetry.  This prediction has been
approved in computer simulations for
hard-disc (short-range) interactions and Coulomb (long-range) potential
\cite{Shklovskii}, so it is probably quite general and valid for all kinds
of isotropic
pair interactions.

In this letter we demonstrate a macroscopic realization of 2D
crystals, which allows to check experimentally the relation between
crystal energy and symmetry.  We use an array of permanent
magnetic particles floating on the surface
of the water \cite{Flanders,Rose-Innes,we}.  This array is confined
by a nonuniform magnetic field of a circular coil
with nearly parabolic distance dependence
$B=B_{0}(1+r^{2}/D^{2})$ \cite{we}.  Here, $B_{0}$ is the field in the
center, and $D$ is a parameter (Fig.1).  Similar systems consisting
of superparamagnetic colloidal particles on the liquid surface and in
the presence of  uniform magnetic field have been studied
in the context of 2D-melting\cite{Zahn,bubeck} and pattern formation
\cite{takahashi}.  In contrast to these works, we use macroscopic
particles for which the corresponding plasma parameter
$\Gamma=E_{int}/k_{B}T\approx 10^{13}*$ is enormous.  Thus, the
Brownian motion is insignificant.

We study 2D cluster configurations vs particle number $N$.  We add/remove one
particle at the cluster boundary in the presence of the constant external
field and wait for
equilibration.  To accelerate equilibration, we stir the particles
 using an oscillating magnetic field (Fig.1) with slowly decreasing
amplitude (stimulated annealing).  After several minutes the particles
self-assemble into a well-ordered cluster with quite reproducible
configuration which remains stable for many days.

An equilibrium particle configuration is a hexagonally packed circular
cluster (Fig.1, lower panel) with pronounced shell structure.  In all
clusters there are six ``topological'' disclinations \cite{Shklovskii} that
reside close to the boundary.  This resolves the conflict between
circularly symmetric magnetic confinement and hexagonal packing,
arising from $1/r^{n}$ interaction potential\cite{hexagonal}.  In
addition, many clusters (and all clusters with $N>120$) have few
dislocations \cite{Seshadri,Shklovskii} which reside at the boundary or are
attached to the topological disclinations.  There are highly
symmetrical clusters with pronounced facets, which are usually free of
dislocations.  They occur at ``magic'' numbers
(1,7,19,37,55,85\ldots) which roughly
correspond to the number of particles in a perfect hexagon, namely,
$3s(s+1)+1$ where $s=0,1,2...$.

Cluster configuration is determined by the interaction energy
$E_{int}$ (dipole-dipole repulsion) and by the radially-dependent part
of the confinement energy $E_{field}$, as follows \cite{we}:
\begin{equation}
E_{int}=\frac{\mu_{0}}{4\pi}*\sum_{i<j}\frac{m_{i}m_{j}}{|r_{i}-r_{j}|^3};
E_{field}\approx
\frac{B_{0}}{D^{2}}\sum_{i}{m_{i}r_{i}^{2}}.\label{energy}
\end{equation} 
Here $r_{i}$ and $m_{i}$ are the particle position and magnetic moment.
For identical particles with magnetic moment $m$ we introduce energy
and distance scales:
\begin{equation}
E_{0}=\left(\frac{\mu_{0}m^{2}}{4\pi}\right)^{\frac{2}{5}}\left(\frac{B_{0}m}{D^
{2}}\right)^{\frac{3}{5}};
r_{0}=\left(\frac{\mu_{0}mD^{2}}{4\pi B_{0}}\right)^{\frac{1}{5}}\label{scales}
\end{equation} 
and recast the total energy  in dimensionless form as:
\begin{equation}
E=E_{int}+E_{field}=\sum_{i<j}\frac{1}{|r_{i}-r_{j}|^3}
+\sum_{i}r_{i}^{2}\label{dimensionless}
\end{equation} 
Generally speaking, Eq.\ref{dimensionless} allows to find $E$ in
units of $E_{0}$ (Eq.\ref{scales}) by summation of $\sim N^{2}$
terms using the measured particle positions. However, this may be
reduced to the sum of only $N$ terms due to special relation
between $E_{int}$ and $E_{field}$ in the equilibrium state.  Indeed,
let's consider an energy (Eq.\ref{dimensionless}) variation
upon small uniform deformation
$r_{i}\rightarrow r_{i}(1+u)$, where $u<<1$. We find
\begin{equation}
\delta E=-u(3E_{int}-2E_{field})+u^{2}(6E_{int}+E_{field})\label{quadratic}
\end{equation}
In the equilibrium state the linear term  vanishes which implies
$E_{int}:E_{field}=2:3$. Therefore,
\begin{equation}
E=\frac{5}{3}E_{field}=\frac{5}{3}I\label{relation}
\end{equation} 
Here, $I=\sum_{i}r_{i}^{2}$ is the moment of inertia. It contains
only $N$ terms and is easily measured. Quadratic term
in Eq.\ref{quadratic} yields bulk compression modulus
$K=S^{-1}\partial^{2}E/\partial u^{2}=6E/S$, where $S$ is cluster area.

For each cluster with particle number $N$  we take an image, using
computer-controlled CCD camera, and
determine the particle positions, using MATLAB. Then, we calculate the
moment of inertia
$I=\sum_{i}r_{i}^{2}$, area $S$, and the
center of mass position.  The total energy $E$ is calculated through
Eq.\ref{relation} with the accuracy of 0.15\% as found in
experiment with the same cluster which was shaked between the
measurements and then allowed to relax.  The scale $r_{0}$ (Eq.\ref{scales})
is found from the measurement of particle positions in the clusters
consisting of 2-7 particles, whose configuration is evident.  Figure 2
shows $S(N),E(N)$, and the chemical potential $\mu(N)=\partial
E/\partial N$. These parameters increase with $N$ according to a power-law
dependence $\sim N^{3/5}$ which may be explained within a continuum
approximation.  To this end we calculate total energy in the limit of
large $N$ by replacing the sum (Eq.\ref{dimensionless}) by
the integral and assuming constant density $\rho$.  We find
\begin{equation}
E\approx N\int_{2a}^{\infty}\frac{\pi r\rho dr}{r^3}+\frac{NR^{2}}{2}=
\frac{N}{2}\left({1/a^{3}}+R^{2}*\right)\label{epsilon}
\end{equation} 
where $R$ is the cluster radius, $2a$ is the lattice constant, and
$\rho\approx 1/\pi a^{2}*$.
The first term in
Eq.\ref{epsilon} represents $E_{int}$, while the second term
represents $E_{field}$.  Since according to Eq.\ref{quadratic} their ratio
is 2:3, Eq.\ref{epsilon} yields $2/a^{3}=3R^{2}$.
Combining it with the obvious relation $R\approx aN^{\frac{1}{2}}$, we find
$R=\left(\frac{3}{2}\right)^{\frac{1}{5}} N^{\frac{3}{10}}$,
$a =\left(\frac{3}{2N}\right)^{\frac{1}{5}}$, and a single particle energy
$\epsilon=E/N=\frac{5}{4a^{3}}$, so that
\begin{equation}
\epsilon \sim 0.98N^\frac{3}{5};
\mu\sim 1.57N^\frac{3}{5};
S\sim 3.69N^\frac{3}{5};
K\sim 1.59N\label{3/5}
\end{equation}

Figure 2 shows that Eq.\ref{3/5} accounts fairly well for the
experimentally found $N^{3/5}$ dependence for
$S,\mu,\epsilon$ (the prefactors are by 20\% smaller and this will be
discussed below) and for the linear dependence of $K$.  Deviations at
small $N$ result from the surface energy and are accounted by a more
advanced model (to be discussed elsewhere).

Note deviations of $S,\mu, \epsilon, K$ from smooth dependences
(Fig.2) which look like random fluctuations although they are quite
reproducible and exceed the experimental errors by five times.  These
fluctuations appear in all cluster properties
and are very prominent in $\mu$, where they
amount up to 15\%.  Note that there are regions in which the chemical
potential remains flat or even decreases with addition of new
particles (Fig.2).

Figure 3 shows expanded view of the fluctuations in total energy
$\delta E$ (after subtraction of a smooth background). It also shows
second derivative of the total energy, $\frac{\partial^{2}* E}{\partial
N^{2}}$.
In the context of quantum
dots the latter corresponds to charging energy.  Pronounced positive
peaks in $\frac{\partial^{2}* E}{\partial N^{2}}$ occur for "magic"
clusters which have almost perfect symmetry.  The
energy of these clusters is lower (note dips in $\delta E$ at Fig.  3)
while the density and the bulk compression modulus is higher
(not shown here).  The magnitude of the fluctuations is $\delta E
\approx 0.1\epsilon$ and $\delta S\sim S/N$.  In three experimental
runs (one for decreasing $N$ and two for increasing $N$) the energy
dips occur
almost at the same magic numbers (7,19,35,55,86,118..).   While magic
numbers are well-known for
free-standing 3D clusters \cite{Martin}, and almost the same numbers have
been found in computer
simulations for 2D clusters \cite{Bedanov,Lozovik,Shklovskii}, we are
unaware of previous experimental observations of magic 2D
clusters in lateral confining field.

Koulakov and Shklovskii \cite{Shklovskii} have recently related magic
numbers and energy fluctuations to the cluster symmetry and to the
center of mass position in the crystalline lattice.  While the
absolute position of the center of mass is always at the
minimum of the external potential, its position with respect to the
unit cell is not fixed.  Figure 4 shows a unit cell of the hexagonal
lattice where (A), (B) and (C) represent points of high symmetry
(six-fold, three-fold and two-fold, respectively).
Ref.\cite{Shklovskii} shows that the energy minimum is achieved when
the center of mass of the whole cluster is located at one of these points.
We measured
distribution of the center of mass positions in the unit cell of our
2D dipole clusters and indeed found that it is strongly peaked
at (A), (B) or (C) (Fig.4) .  A lower panel in Fig.3 also shows $N_{cm}$, a
number of nearest neighbors at the center of mass position vs
particle number.  [We take
one of the points (A), (B), or (C) which is the closest to the center
of mass and assign $N_{cm}=6$ for (A), $N_{cm}=3$ for (B), and
$N_{cm}=2$ for (C)].  $N_{cm}$ changes
quasiperiodically in $N^{\frac{1}{2}}$. There are continuous
ranges where the center of mass resides in the position of six-fold
symmetry (A).  In these ranges the dips of energy and pronounced
peaks of $\frac{\partial^{2}* E}{\partial N^{2}}$ occur.  Shallow dips
correspond to position (B).
Ref.\cite{Shklovskii} shows that upon addition of new particles, the
cluster undergoes energetically expensive elastic and plastic
(formation of dislocations) deformations in order to keep the center
of mass at the position of high symmetry.  This occurs until another
position of high symmetry becomes energetically favorable.  Then, upon
addition of one more particle, an avalanche occurs.  This means
rearrangement of many particles, whereby the center of mass moves to a
large distance $\sim a$.  The avalanches take place with periodicity
of $\sim N^{\frac{1}{2}}$ which corresponds to addition of a new
crystalline row.  In Fig.4 we present the distance between the center of mass
and the nearest particle  and indeed observe
that it does not vary gradually, but in a step-like fashion.

We estimate the magnitude of fluctuations as follows.  Assume a
perfectly symmetric cluster.  Addition of new particles does not lead
to the center of mass displacement until approximately one crystalline
row or $\Delta N \sim (N/3)^{1/2}$ particles are added.  If the cluster were
incompressible, corresponding area increase would be $\delta S\sim S
\Delta N/N$.  Since our clusters are compressible, immobility of the
center of mass means that the cluster adopts new particles without
appreciable change in area.  Therefore, deviation from the smooth
dependence $S(N)$ is $\delta S \sim S/(3N)^{1/2}$ and it arises from
cluster compressibility and deformation.  If the deformation were
purely elastic, then $u=\delta S /2S$ and the corresponding energy change
would be $\delta E=KSu^{2}/2\sim 0.25\epsilon$.  Experimentally observed
energy fluctuations ($0.1\epsilon-0.05\epsilon$) are smaller due to
formation of the dislocations that lead to relaxation of elastic
stresses.

Our experimental results should be compared to numerical simulations
of Ref.\cite{Lozovik} for 2D dipole clusters ($N<80$) in a parabolic
confinement.  Experimentally found shell structures for $N<37$ are
almost identical to those found in simulations, while at $37<N<80$
they differ in half of the cases (the difference is mostly in the
configuration of outer shells) and are more close to those found in
numerical simulations for Coulomb clusters\cite{Bedanov}.  Energy
fluctuations found in experiment and in Ref.\cite{Lozovik} occur
almost at the same ``magic'' numbers, although the experimentally
found magnitude of $\delta E$ is four times bigger than that found in
simulations\cite{Lozovik}. The single particle energies  at small $N$
almost do not differ, while at large $N$ the experimental values are
lower than those found in numerical simulations (see
Fig.3). We find empirical dependence,
$\epsilon_{exp}/\epsilon_{sim}\sim 1-0.1\log N$ in the whole range of
$N$.  This small difference between $\epsilon_{exp}$ and $\epsilon_{sim}$
may be attributed to (i) additional attractive capillary forces
between the particles which
logarithmically decay with distance \cite{takahashi}; (ii)
deviation of the confining field from parabolicity which appears
mostly at large $N$ when the cluster radius becomes comparable to the
radius of the coil. Both these factors lead to the cluster compression
and to the underestimate of the total energy using Eq.\ref{relation}.

This work was supported by the Israeli Ministry of Science and Arts
(through the Israeli-French collaboration) and VW foundation.
We are grateful to B. Laikhtman and A.A.  Koulakov for valuable
discussions.


\pagebreak
\section*{Figure captions}
\begin{enumerate}
\item [Fig.1] Experimental setup.  Small rare-earth permanent magnets
encapsulated within styrofoam discs are floating on the surface of the
water and repel each other.  They are confined within a nonuniform
magnetic field produced by external coil.  Container diameter is 100
cm, magnet size is 3 mm.  Lower panel shows two cluster images with
superimposed Delaunay triangulation.  Gray circles stand for normal
coordination number ($Z=6$), filled circles stand for $Z=5$ (positive
disclination), and open circles stand for $Z=7$ (negative
disclination).  $N$=85 - a``magic'' cluster.  Note hexagonal structure
and six topological disclinations at the boundary.  $N$=70- an
ordinary cluster.  Note a dislocation (5-7 pair) attached to a
disclination.
\item[Fig.2] Single particle energy $\epsilon$, the chemical potential
$\mu$ and cluster area $S$ vs particle number $N$ (in dimensionless units).
Continuous lines show experimental data (smoothed); dashed lines show
asymptotic
$N^{3/5}$dependence; dashed-dotted line shows  results of numerical
simulations of Ref.\cite{Lozovik} (multiplied by
0.8) for the energy of 2D dipole clusters in parabolic confinement.
The inset shows bulk compression modulus $K$.
\item[Fig.3] Fluctuations of the total cluster energy $\delta E$ and
``charging energy'' $\frac{\partial^{2} E}{\partial N^{2}}$ (note
inverted scale) vs particle number $N$.  The lowest curve shows number of
nearest neighbors, $N_{cm}$, to the center of mass
position.  Note correlation between three curves.  Big
positive values of $\frac{\partial^{2}* E}{\partial N^{2}}$ (dips in
the figure) correspond to dips in $\delta E$.  Deep dips occur when
$N_{cm}=6$ and shallow dips correspond to $N_{cm}=3$. ``Magic''
numbers indicated on the Figure correspond to energy dips and almost
perfect clusters.
\item[Fig.4] Center of mass position.  (a) Unit cell and the positions
of high symmetry, (A),(B), and (C).  (b) Distribution of the center of
mass position in the unit cell.  Each point corresponds to a cluster
with a certain particle number $N$.  Note sticking of the center of
mass to positions (A),(B), and (C).  Lower panel shows the distance
between the center of mass and the nearest particle.  Note
$N^{\frac{1}{2}}$ periodicity.
\end {enumerate}
\end{document}